# Collapse helps to probe the structure of particles

Rui Qi
Institute of Electronics, Chinese Academy of Sciences
17 Zhongguancun Rd., Beijing, China
E-mail: rg@mail.ie.ac.cn

We present a possible method to probe the inner structure of particles based on one kind of promising dynamical collapse theory. It is shown that the present decay data of $K_L$ meson indicates that quarks have no inner structure.

## Introduction

As we know, the usual way to probe the inner structure of particles is to use high-energy accelerator. It requires larger and larger energy if we want to probe the smaller and smaller structure of particles. This severely restricts the development of particles physics. For example, present experiments can't determine whether the quarks have inner structure yet. In this short paper, we will present a possible new method to probe the inner structure of particles based on one kind of promising dynamical collapse theory. It is demonstrated that the present decay data of $K_L$ meson may have indicated that quarks have no inner structure.

## A dynamical collapse theory

As to the evolution of the wave function during quantum measurement, present quantum theory provides by no means a complete description. The projection postulate is just a makeshift, while the concrete dynamical process of the projection is undoubtedly one of the most important unsettled problems in quantum theory. Recently the resulting dynamical collapse theory or revised quantum dynamics[1-10] are deeply studied, in which the linear evolution equation of the wave function is replaced by stochastic linear or nonlinear equation.

Here we will briefly introduce one kind of promising dynamical collapse theory[3][5-10], in which the dynamical collapse is driven by the energy difference among the sub-states in the superposition states. As an example, we analyze the evolution of a simple two-level system using the theory. Let the initial wave function of the particle be $\psi(x,0) = a(0)^{1/2} \psi_1(x) + b(0)^{1/2} \psi_2(x)$, which is a superposition of two static states with different energy levels $E_1$ and $E_2$. According to the theory, Since the linear item in the evolution equation of wave function only results in a phase factor, and doesn't influence the collapse results, we only consider the nonlinear stochastic item. Through some mathematical calculations we can work out the density matrix of the two-level system:

$$r_{11}(t) = a(0) \quad \text{-----(1)}$$

$$r_{12}(t) \cong [1 - \frac{(\Delta E)^2}{k \cdot \hbar E_p} t] \sqrt{a(0) b(0)} \quad \text{------(2)}$$

$$r_{21}(t) \cong [1 - \frac{(\Delta E)^2}{k \cdot \hbar E_p} t] \sqrt{a(0) b(0)} \quad \text{------(3)}$$

$$r_{22}(t) = b(0) \quad \text{------(4)}$$

where $k$ is a dimensionless constant, $\hbar$ is the Planck constant h divided by $2\pi$, $E_p$ is Planck energy, $\Delta E = E_2 - E_1$, is the difference of the energy between these two states in the superposition. The collapse time for such two-level system is: $t_c \approx k \frac{\hbar E_p}{(\Delta E)^2}$. It can be seen that the energy difference is larger, the collapse time is shorter. Besides, it should be denoted that, as to the superposition state of many particles such as $\psi(x,0) = a(0)^{1/2} \psi_1(x) \varphi_1(x) + b(0)^{1/2} \psi_2(x) \varphi_2(x)$, the whole energy difference is the sum of the absolute values of the energy difference of every particle.

## The method to probe the structure of particles

Now we will demonstrate how the collapse may help to probe the inner structure of particles. According to the dynamical collapse theory, the collapse time will be essentially relevant to the structure of particles in the superposition state. The general principle is that the particle has deeper inner structure, the whole energy difference of the superposition state of the particle is larger, and the collapse time is shorter.

Here we present a simple example to explain this principle. We assume the initial state of a particle is the superposition state of two particles 1 and 2 $|1> + |2>$, the particles 1 and 2 have the same mass; If the particles 1 and 2 have no further structure, then the superposition state will not collapse since the energy difference between the states in the superposition is zero. But if the particles 1 and 2 have some kind of inner structure, for example, they are composed of two sub-particles respectively, then the conclusion will be essentially different. We assume the particle 1 is composed of particles 11 and 12, and the particle 2 is composed of particles 21 and 22, in which these sub-particles possess different masses. Now the initial state turns to be $|11>|12> + |21>|22>$. The whole energy difference influencing the dynamical collapse is $E_1 = |E_{11} - E_{12}| + |E_{12} - E_{22}|$, which is not zero. Then the initial state will collapse, and the collapse time is in the level $t_c \approx \dfrac{\hbar E_p}{(\Delta E)^2}$. Furthermore, if the particles 11 etc are also composed of smaller particles, then it can be seen that the whole energy difference will generally turn to be larger, and the collapse time will be shorter. Thus the collapse time of such superposition state will tell us some information about the possible inner structure of the particle. This conclusion essentially relies on the above promising dynamical collapse theory.

But one question is left, i.e. where to find the above seemingly bizarre superposition state? Fortunately, the $K_L$ meson just has one. The state of $K_L$ meson can be written as follows:

$$|K_L> = |K_O> + |\overline{K}_O> = \frac{1}{\sqrt{2}}[|s>|\overline{d}> - |d>|\overline{s}>] \quad \text{------ (5)}$$

This is just a superposition state of two particles $K_O$ and $\overline{K}_O$, which has the same mass and is composed of two quarks respectively. According to the above dynamical collapse theory, if the particles $K_O$ and $\overline{K}_O$ have no inner structure, then the whole energy difference is zero, and the superposition state will not collapse. But if the particles $K_O$ and $\overline{K}_O$ is composed of two quarks as shown above, then the whole energy difference will be approximately 400Mev, and the corresponding collapse time will be in the level of $10^{-4}$ s.

Now there appears a very interesting question, i.e. do the quarks have inner structure? Our method and the decay data of $K_L$ meson may provide a possible answer. As Fivel had denoted[6], the above collapse time surprisingly coincides with the decay data of $K_L$ meson. This strongly suggests that the dynamical collapse of the state of $K_L$ meson indeed happens, and the collapse time is also the same as that predicted according to the quarks structure and dynamical collapse theory. Then we may get three very useful conclusions. One is that if the prediction of the dynamical collapse theory is right, then the quarks will possess no inner structure, and be the real basic particles. If the quarks is composed of other sub-particles, then the dynamical collapse theory will generally give a different prediction of the collapse time, which is not consistent with the decay data. The second is that if the quarks theory is right, then the decay data will provide the first confirmation of the dynamical collapse theory. The last is that, as Fivel suggested[6], the collapse of wave function may provide an economical explanation of CP violations, namely CP violations completely results from the collapse of wave function.

## Conclusions

In this paper, we present a possible method to probe the inner structure of particles based on one kind of promising dynamical collapse theory. The general principle is that the particle has deeper inner structure, the whole energy difference of the superposition state of the particle is larger, and the collapse time is shorter. It is also shown that the present decay data of $K_L$ meson may indicate that quarks have no inner structure.